\documentclass[preprint,showpacs,preprintnumbers,amsmath,amssymb]{revtex4}
\usepackage{graphicx}
\usepackage{dcolumn}
\usepackage{bm}
\begin{document}
\preprint{APS/123-QED}
\title{$^{63}$Cu, $^{35}$Cl, and $^{1}$H NMR
in the
$S=\frac{1}{2}$ Kagom\'e Lattice ZnCu$_{3}$(OH)$_{6}$Cl$_{2}$}
\author{T. Imai$^{1,2}$, E. A. Nytko$^{3}$, B.M. Bartlett$^{3}$, M.P. 
Shores$^{3}$, and D. G. Nocera$^{3}$}
\affiliation{$^{1}$Department of Physics and Astronomy, McMaster University, Hamilton, ON L8S4M1, Canada}
\affiliation{$^{2}$Canadian Institute for Advanced Research, Toronto, Ontario M5G1Z8, Canada}
\affiliation{ $^{3}$Department of Chemistry, M.I.T., Cambridge, Massachusetts 02139}
\date{\today}
\begin{abstract}
ZnCu$_{3}$(OH)$_{6}$Cl$_{2}$ ($S=\frac{1}{2}$) is a promising new candidate for an ideal Kagom\'e Heisenberg antiferromagnet, because there is no magnetic phase transition down to $\sim$50~mK.  We investigated its local magnetic and lattice environments with NMR techniques.  We demonstrate that the intrinsic local spin susceptibility {\it decreases} toward $T=0$, but that slow freezing  of the lattice near $\sim$50~K, presumably associated with OH bonds, contributes to a large increase of local spin susceptibility and its distribution.  Spin dynamics near $T=0$ obey a power-law behavior in high magnetic fields.
\end{abstract}
\maketitle
A major challenge in condensed matter physics today
is identifying a model material for investigating {\it spin
liquid} \cite{Anderson, PhysicsToday}.  Searching for exotic electronic states
without magnetic long range order, such as Kagom\'e Heisenberg antiferromagnets, constitutes a common thread in a
wide range of research fields, from high temperature
superconductivity to low dimensional quantum magnetism.   
Over the last decade, many candidate materials have been investigated as model systems for a Kagom\'e lattice \cite{SrCr,Jarosite,CuV,Review}.
However, they mostly exhibit a magnetically ordered or
spin-glass-like state at low temperatures. 
A recent breakthrough in the hunt for a spin liquid state\cite{PhysicsToday}
is the successful synthesis \cite{Shores} and characterization
\cite{Helton} of ZnCu$_{3}$(OH)$_{6}$Cl$_{2}$ (herbertsmithite), a chemically pure spin $S=\frac{1}{2}$ Kagom\'e lattice.
As shown in Fig.\ 1, three Cu$^{2+}$ ions form a triangle, and a
network of  corner-shared triangles form a Kagom\'e lattice.
The $S=\frac{1}{2}$ spins on Cu sites are mutually frustrated by
antiferromagnetic super-exchange interaction $J\sim 170$ K
\cite{Helton,Singh1}, hence the possibility of a spin liquid ground state.

Recent  measurements of ZnCu$_{3}$(OH)$_{6}$Cl$_{2}$ with 
bulk magnetic susceptibility, $\chi_{bulk}$\cite{Helton},
specific heat\cite{Helton}, neutron scattering on
powders\cite{Helton}, $\mu$SR\cite{Ofer,Mendels}, and $^{35}$Cl NMR\cite{Ofer} have
established that ZnCu$_{3}$(OH)$_{6}$Cl$_{2}$ remains paramagnetic
down to at least $\sim50~mK$ with no evidence of magnetic long range order. These findings indeed point towards the possible realization of a frustrated spin liquid state with the Kagom\'e symmetry.  
However, very little is known beyond the paramagnetic nature of the ground state.  For example, the bulk averaged susceptibility, $\chi_{bulk}$, reveals a mysterious sharp
{\it increase} below $\sim 50$ K \cite{Helton}.  This clearly contradicts the
predictions of various theoretical calculations: series expansions
predict a {\it decrease} of $\chi_{bulk}$ below $T\sim J/6$ with a
gap\cite{Elstner,Mila}, while the recent Dirac Fermion model
predicts linear behavior in $T$ towards $T=0$ \cite{Lee}.  Does this apparent contradiction mean that ZnCu$_{3}$(OH)$_{6}$Cl$_{2}$ is not a good Kagom\'e model system after all, or that extrinsic effects other than Kagom\'e Heisenberg interaction, such as mixing of Zn ($S=0$) into Cu ($S=\frac{1}{2}$) sites\cite{Vries,Bert} and Dzyaloshinsky-Moriya (DM) interactions\cite{Singh1,Singh2}, simply mask the intrinsic Kagom\'e behavior below $\sim 50~K$?  What about spin dynamics?  Do spin fluctuations slow down toward a critical point, or are they gapped \cite{Elstner,Mila,Lee}?

In this {\it Letter}, we report a $^{63}$Cu, $^{35}$Cl, and $^{1}$H NMR
investigation of ZnCu$_{3}$(OH)$_{6}$Cl$_{2}$ for a broad range of magnetic fields and frequencies. 
Taking full advantage of the local nature of NMR techniques, 
we uncover hitherto unknown properties of
ZnCu$_{3}$(OH)$_{6}$Cl$_{2}$.  First, from the observation of the broadening of $^{35}$Cl NMR lineshapes, we will demonstrate that local spin susceptibility, $\chi_{loc}$, has a large distribution throughout the sample.  Moreover, the smallest components of $\chi_{loc}$ actually {\it saturate} and even  {\it decrease} 
with $T$ below $T\sim 0.2J$, even though the bulk averaged $\chi_{bulk}$ increases as $\sim 1/T$.  The observed decrease of  $\chi_{loc}$ is precisely what the intrinsic spin susceptibility of a Kagom\'e Heisenberg antiferromagnet is expected to show.  Second, from the comparison of $^{1}$H and $^{35}$Cl
nuclear spin-lattice relaxation rates, $^{1,35}(1/T_{1})$, we present unambiguous evidence for slow freezing of the lattice near $\sim50~K$, most likely due to orientational disorder of OH bonds.  We suggest that this subtle freezing of lattice distortion enhances the DM interactions, and is key to understanding the aforementioned upturn of bulk-averaged spin susceptibility, $\chi_{bulk}$, below $\sim50~K$ that masks the intrinsic Kagom\'e behavior of $\chi_{loc}$.  Third, we demonstrate from the measurements of 
the $^{63}$Cu nuclear spin-lattice relaxation rate, $^{63}(1/T_{1})$, that in the presence of a high magnetic field the low 
frequency Cu spin fluctuations {\it grow} without a gap below $\sim 30$ K satisfying a simple power-law.

In Fig.\ 2, we show representative $^{35}$Cl NMR lineshapes.  For
these measurements, we cured a powder sample in glue in a magnetic field
of 9 Tesla.  From powder x-ray diffraction measurements, we confirmed 
that approximately 20\% of the powder is uniaxially aligned along the
c-axis.  In fact, we observe a sharp c-axis central peak near 35.02 MHz
(marked as {\it B//c} in Fig.\ 2) arising from particles
oriented along the c-axis.  The "double horns" marked as \#1 and \#2 are split by the nuclear quadrupole interaction, and arise from the randomly oriented portion of the powder (i.e. 80\% of the sample) \cite{Ofer}.   Notice that the whole $^{35}$Cl NMR lineshape begins to
tail-off toward lower frequencies below $\sim$50~K.  The resonance frequency of the sharp c-axis central peak and its distribution depends on the NMR Knight shift, $^{35}K$, induced by $\chi_{loc}$.  Hence the observed line broadening implies that {\it $\chi_{loc}$ varies depending on the location within the sample} below $\sim 50~K$.

In Fig.\ 3, we summarize the $^{35}$Cl NMR Knight shifts $^{35}K$ and $^{35}K_{1/2}$
 deduced from the lineshapes, together with $\chi_{bulk}$ as observed by SQUID.  $^{35}K$ corresponds to the central peak above $\sim$ 45~K as determined by FFT techniques. Below $\sim$ 45~K, where the central peak is smeared out by line broadening, we determined $^{35}K$ as the higher frequency edge of the central peak from point-by-point measurements,   {\it i.e.} $^{35}K$ represents the smallest component of the distributed $\chi_{loc}$.   $^{35}K_{1/2}$ corresponds to the 
half-intensity position of the central peak on the lower frequency side of the spectrum.  Quite generally, $^{35}K = A_{hf}\chi_{local} + ^{35}K_{chem}$, where $A_{hf}$ is the magnetic hyperfine interaction between
$^{35}$Cl nuclear spin and nearby Cu electron spins, and
$^{35}K_{chem}$ is a very small, temperature independent chemical
shift.  In the present case, from the comparison with $\chi_{bulk}$,
we can estimate $A_{hf}\sim -3.7 \pm 0.7$ kOe/$\mu_{B}$.  The negative
sign of $A_{hf}$ makes the overall sign of $^{35}K$ negative.
	Accordingly, we have inverted the vertical scale of Fig.\ 3.

	We wish to comment on two important aspects of Fig.\ 3.  First,
$^{35}K$ follows Curie-Weiss behavior all
the way from 295~K down to $\sim$25~K.  This clearly differs from
$\chi_{bulk}$ which begins to deviate from Curie-Weiss behavior
below temperatures as high as $\sim$150~K \cite{Helton}.  On the other hand, series-expansion calculations
indicate that the Kagom\'e lattice follows Curie-Weiss behavior down
to $T\sim J/6\sim 25$ K \cite{Elstner}. Our Knight shift data
demonstrate that $\chi_{local}$ of some ideal segments of the Cu$^{2+}$ Kagom\'e
lattice in this material indeed show such behavior. Theoretical models also predict
that below $T\sim J/6\sim 25$~K, $\chi_{bulk}$ begins to decrease
exponentially with a small gap\cite{Elstner}, or linearly\cite{Lee}.
As shown in Fig.\ 2, the $^{35}$Cl NMR lineshape begins to transfer
some spectral weight to higher frequencies below 25~K. Recalling that
$A_{hf}$ is negative, {\it this is consistent with vanishing spin
susceptibility, $\chi_{local}$, near $T=0$ for some parts of the
Kagom\'e lattice}.  After the initial submission of this {\it Letter}, Olariu {\it et al.} also observed a similar decrease of $^{17}$O NMR Knight shift and arrived at the same conclusion \cite{Olariu}. 

Another important aspect of Fig.\ 3 is that $^{35}K_{1/2}$ begins
to deviate from the aforementioned Curie-Weiss fit in a manner
similar to $\chi_{bulk}$, as the $^{35}$Cl NMR line
gradually broadens to lower frequencies. It is important to realize that
$^{35}(1/T_{1})$ also begins to increase in the same temperature range below $\sim$150~K (see Fig.\ 4).
Below 50~K, where $^{35}(1/T_{1})$ shows a peak, the $^{35}$Cl NMR
line shows a dramatic broadening to lower frequencies.  Fig.\ 2 and Fig.\ 3 establish that
$^{35}K_{1/2}$ follows the same trend as $\chi_{bulk}$, i.e. {\it some segments of the Kagom\'e
lattice have large and distributed local spin susceptibility
$\chi_{local}$}, and {their temperature dependence is different from
the smaller $\chi_{local}$ as represented by $^{35}K$.}   $\chi_{bulk}$ simply represents a
bulk average of $\chi_{local}$.

In passing, we recall that earlier $\mu$SR Knight shift $K_{\mu SR}$
measurements by Ofer {\it et al.} \cite{Ofer} showed identical behavior
between $K_{\mu SR}$ and $\chi_{bulk}$. They concluded that the
upturn of $\chi_{bulk}$ below 50 K is not caused by impurity spins
but is a bulk phenomenon. Our new results in Fig.\ 3 do not contradict
these  $\mu$SR data.  $K_{\mu SR}$ was
deduced by assuming a Gaussian distribution of $\chi_{local}$, hence
by default $K_{\mu SR}$ represents the central value of the presumed
Gaussian distribution.  That explains why $K_{\mu SR}$ shows
behavior similar to $\chi_{bulk}$ and $^{35}K_{1/2}$.

Next, we turn our attention to the dynamics of lattice and spin degrees of freedom.  Fig.\ 4 shows the
temperature dependence of the $^{35}$Cl nuclear spin-lattice relaxation
rate, $^{35}(1/T_{1})$, measured at the central peak frequency in various magnetic fields, $B$.  We also plot 
$^{1}(1/T_{1})$ for $^1$H NMR in 0.9 Tesla, and $^{63}(1/T_{1})$ for $^{63}$Cu NMR in 8 Tesla.  We have overlaid 
$^{1,63}(1/T_{1})$ on $^{35}(1/T_{1})$ measured in comparable magnetic fields by scaling the vertical axis.

We can draw a number of conclusions from Fig.\ 4.  First, let us focus on $T$ and $B$ independent results of $^{35}(1/T_{1})$
above $\sim$150~K.  This high temperature regime is easily understandable within Moriya's theory for the exchange narrowing limit of Heisenberg antiferromagnets, where we should expect $^{35}(1/T_{1})_{exc}\sim A_{hf}^{2}/J = constant$ for $T>J\sim 170$ \cite{MoriyaExc}.  If we assume $J\sim170$K and $A_{hf}\sim -4$ kOe/$\mu_{B}$, we can estimate $^{35}(1/T_{1})_{exc}\sim 4$ sec$^{-1}$.  This is in excellent agreement with our result.

Another important feature is that $^{35}(1/T_{1})$ begins to increase below $\sim$150~K and peaks near $\sim 50~K$.  Since  $^{35}$Cl is a quadrupolar nucleus with nuclear spin $I=\frac{3}{2}$, the observed enhancement may be caused by slow fluctuations of the lattice via nuclear quadrupole interactions, as well as by Cu spin fluctuations.  To discern the two possibilities, we show $^{1}(1/T_{1})$ of $^1$H measured at 0.9~Tesla for comparison.  The $^{1}(1/T_{1})$ data nicely interpolate $^{35}(1/T_{1})$ in the field-independent regime above $\sim$150~K and $^{35}(1/T_{1})$ measured at a comparable magnetic field (1~Tesla) at low-temperatures, without a peak near $\sim 50~K$.  Since $^1$H has $I=\frac{1}{2}$, $^{1}(1/T_{1})$ has no contributions from lattice fluctuations.  Therefore we conclude that {\it the peak of  $^{35}(1/T_{1})$ near $\sim 50~K$ arises from enhancement of lattice fluctuations at the NMR frequency}.  The peak of  $^{35}(1/T_{1})$ shifts to progressively lower temperatures as we lower the $^{35}$Cl NMR frequency, from $50~K$ (35 MHz at 8.3~Tesla), $46~K$ (18~MHz at 4.4~Tesla) to $40~K$ (10~MHz at 2.4~Tesla).  This means that the typical frequency scale of lattice fluctuations is 35~MHz at 50~K, 18~MHz at 46~K, and 10~MHz at 40~K, and the Kagom\'e lattice becomes static below 40~K.  Given that no structural phase transition has been detected by x-ray and neutron scattering techniques, the observed freezing of the lattice near $\sim 50~K$ must be a very subtle effect.  In fact, our careful measurements of the quadrupole split $\pm \frac{1}{2}$ to $\pm \frac{3}{2}$ satellite transitions of $^{35}$ Cl NMR didn't detect any noticeable changes, either.  All pieces put together, we suggest that the lightest elements in the lattice, {\it i.e.} OH bonds, must be freezing with random orientations, with only subtle effects on heavier atoms.  

Regardless of the exact nature of the freezing of the lattice, our observation provides a major clue to 
understanding the mysterious behaviors of  $\chi_{bulk}$ and $\chi_{local}$.  Recent structural studies revealed that up to 6~\% of Cu sites may be occupied by Zn to create unpaired defect spins \cite{Vries}.  There is no doubt that such antisite disorder would enhance $\chi_{bulk}$ and  contribute to the large distribution of $\chi_{local}$ at low temperatures.  However, it is important to realize that $\chi_{bulk}$ begins to deviate from high temperature Curie-Weiss behavior below $\sim$150~K \cite{Helton}, exactly where $^{35}(1/T_{1})$ begins to grow due to slowing of lattice fluctuations toward $\sim 50~K$.  Furthermore, the crossover between the two different Curie-Weiss behaviors of $\chi_{bulk}$ and $^{35}K_{1/2}$ takes place precisely when the lattice fluctuations die out below $\sim$50~K \cite{Helton}.  Evidently, the anomaly of the lattice correlates well with that of the spins, suggesting it must be playing a major role in the deviation of spin susceptibility from the theoretically expected behavior of ideal Kagom\'e Heisenberg antiferromagnets.  In  fact, recent numerical simulations by Rigol and Singh showed that DM interactions can enhance spin susceptibility below $\sim$50~K \cite{Singh1,Singh2}.  Our experimental finding naturally fits with their theoretical  picture: when OH bonds freeze with random orientation, the hexagonal symmetry of the Kagom\'e lattice slightly breaks down {\it locally}; this would progressively enhance the DM interaction from  $\sim$150~K to  $\sim$50~K, hence leading to an enhancement of $\chi_{bulk}$ and $^{35}K_{1/2}$, as well as a large distribution of $\chi_{local}$.  

Finally, let us focus on the low temperature regime below 20~K, where only spin fluctuations contribute to $(1/T_{1})$.  $^{35}(1/T_{1})$ at $\sim$4~K is independent of magnetic field from $B\sim$0.9~Tesla up to $B\sim$2~Tesla, but larger magnetic fields suppress $^{35}(1/T_{1})$.  This suggests that unpaired paramagnetic spins are fluctuating in low magnetic fields, but applied magnetic field decouples the weak coupling between them.  The exact origin of these paramagnetic spins is not clear, but one obvious possibility is unpaired free spins induced in the near neighbor sites of Zn ($S=0$) ions occupying the Cu sites, i.e. {\it antisite disorder} \cite{Vries,Bert}.  In fact, after the initial submission of the present work, Olariu et al. reported that approximately  30 \% of the integrated intensity of $^{17}$O NMR signals is split off from the main line, which they attributed to the contribution of the 4 nearest neighbor O sites of Zn ions occupying the Cu sites \cite{Olariu}.  Our observation of field dependence is also consistent with a recent report that the application of $B\sim8$~Tesla suppresses defect contributions to dc spin susceptibility in the low $T$ regime \cite{Bert}.  

All the $^{63,35,1}(1/T_{1})$ data measured at various magnetic fields and frequencies decrease towards the zero temperature limit.  That is, we observe no hint of critical slowing down of spin fluctuations towards $T\sim0$, hence any magnetic critical point, if it exists, is still very far from our temperature region.  This is in agreement with an earlier report based on a limited set of  $^{35}$Cl NMR data\cite{Ofer}, but in remarkable contrast with earlier NMR works on frustrated spin systems with Kagom\'e or triangular lattice geometry, where  $(1/T_{1})$ data always show evidence for a magnetic long range order or freezing \cite{Review}.  At $B\sim8$~Tesla,  both $^{35}(1/T_{1})$ and $^{63}(1/T_{1})$ decrease toward $T=0$, obeying a simple power law, $^{63}(1/T_{1}) = T^{\eta}$ with $\eta \sim 0.5$.  Since the slope of the log-log plot of $(1/T_{1})$ increases somewhat at higher fields, $\eta \sim 0.5$ may be somewhat underestimated. We note that spin fluctuation susceptibility is proportional to $^{63}(1/T_{1}T) = T^{\eta-1}$, hence spin fluctuations {\it grow} toward $T=0$.  This is inconsistent with the exponential decrease expected for a gapped Kagom\'e lattice in some theoretical scenarios.  On the other hand, recent field theoretical calculations based on the Dirac Fermion model predicted  a power law, $^{63}(1/T_{1}) = T^{\eta}$, with unknown critical exponent $\eta$ \cite{Lee}.  We note that the issue of power law behavior of spin dynamics is at the focus of the field theoretical approach toward spin liquid systems.  It remains to be seen if the Dirac Fermion model would be consistent with  $\eta \gtrsim 0.5$.   

To summarize, we have presented a site-by-site picture of the new
Kagom\'e material ZnCu$_{3}$(OH)$_{6}$Cl$_{2}$ using NMR techniques.   Our NMR data revealed that both local spin susceptibility and spin dynamics (in high magnetic fields) show aspects that are consistent with theoretical conjectures for ideal Kagome antiferromagnets, despite perturbations from lattice freezing and paramagnetic defects. 

We thank Y.S. Lee, P.A. Lee, S. De'Soto, M. Rigol, R. Singh, and P. Mendels for helpful
communications.  The work at McMaster was supported by NSERC and CIFAR.
  We thank NSF for providing EAN and BMB with predoctoral 
fellowships, and DuPont for providing BMB with a Graduate Fellowship 
Award.

%

\pagebreak

Fig.1\\
 Left : A Kagom\'e lattice.  Right :
Cu$^{2+}$ Kagom\'e layer in ZnCu$_{3}$(OH)$_{6}$Cl$_{2}$.  Cl$^{-}$ site is
above or below the middle of the Cu$^{2+}$ triangles.\\

Fig.2\\
$^{35}$Cl NMR lineshapes  of the $I_{z}=+\frac{1}{2}$ to $-\frac{1}{2}$ central
transition in 8.4
Tesla in a partially ($\sim$20~\%) uniaxially aligned powder sample.  The
sharp peak near 35.02 MHz marked as ``{\it B//c}'' originates from the
particles whose c-axis is aligned along the external
magnetic field. Vertical lines specify the c-axis peak and edge corresponding to $^{35}K$.  Vertical
dashed lines specify the frequency corresponding to $^{35}K_{1/2}$.\\

Fig.3\\
 $^{35}$Cl NMR Knight shift (left scale). 
$^{35}K$ (blue circles).  
$^{35}K_{1/2}$ (red triangles).  The blue solid line
is a fit to Curie-Weiss behavior, $^{35}K =
(22\pm7)/(T-\theta_{CW}) + ^{35}K_{chem}$,  where $\theta_{CW} = 
-240\pm 80$K.  The constant background $^{35}K_{chem}=0.018\%$ is 
shown by a dashed line.
$\chi_{bulk}$  measured by SQUID (dotted line) is also overlaid (right scale).\\

Fig.4\\
Temperature dependence of $^{35}$Cl NMR spin-lattice relaxation rate $^{35}(1/T_{1})$ at various magnetic fields (filled symbols).  Solid line represents a fit to a power law, $^{35}(1/T_{1}) = T^{\eta }$ with $\eta = 0.47$ (8.3 T), 0.44 (4.4 T), 0.2 (2.4 T and 1.0 T).  $^{1}$H relaxation rate in low field (0.9 T), $^{1}(1/T_{1})$, and $^{63}$Cu relaxation rate in high field (8 T), $^{63}(1/T_{1})$, are also superposed on $^{35}(1/T_{1})$ measured in comparable magnetic field. \\

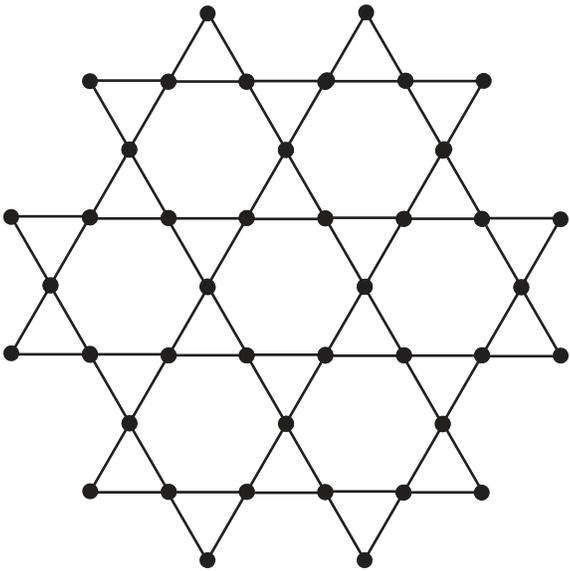
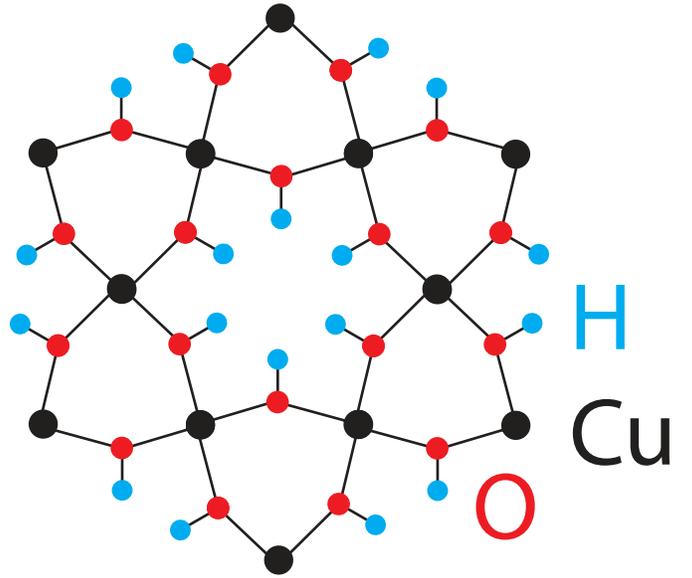
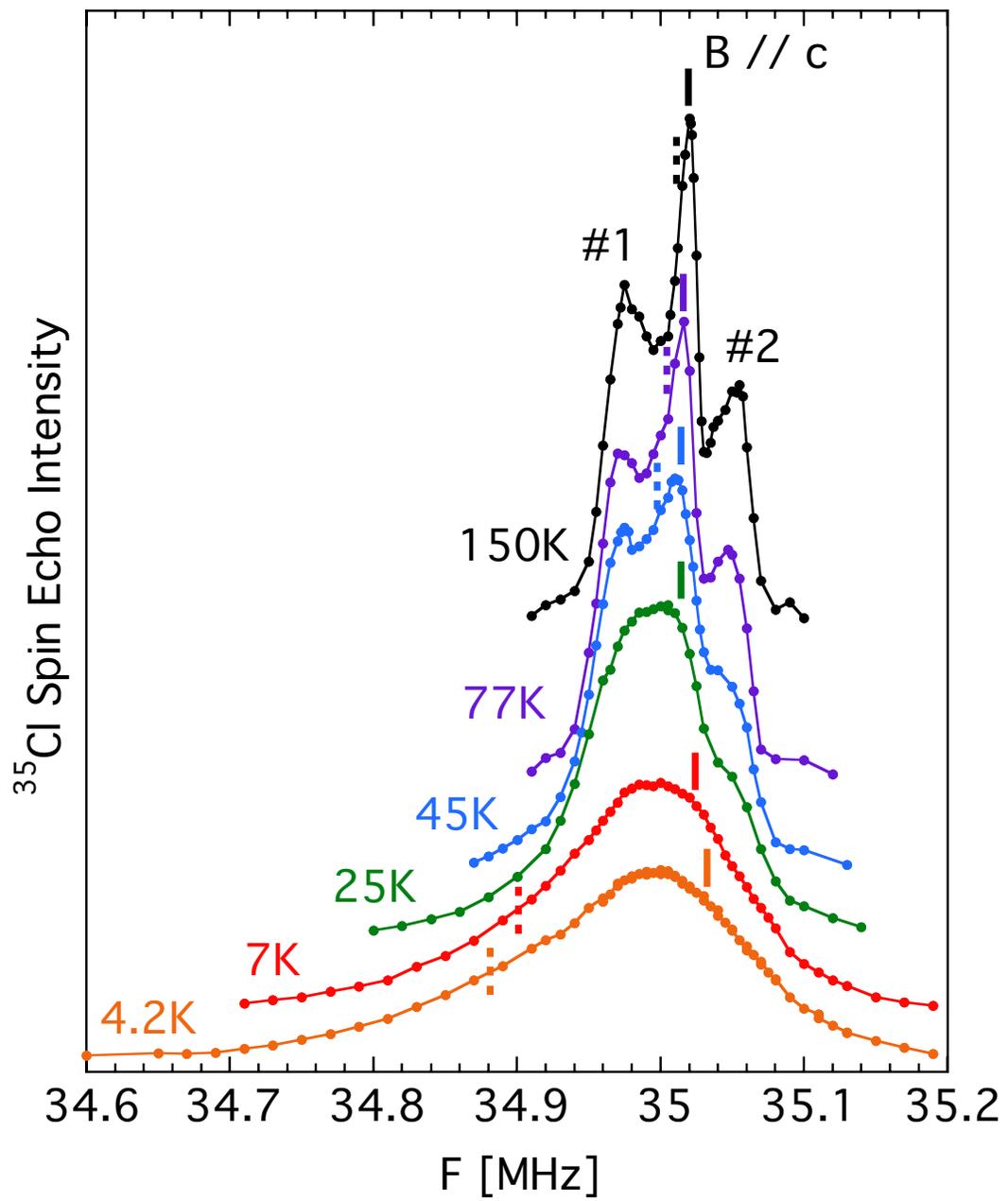
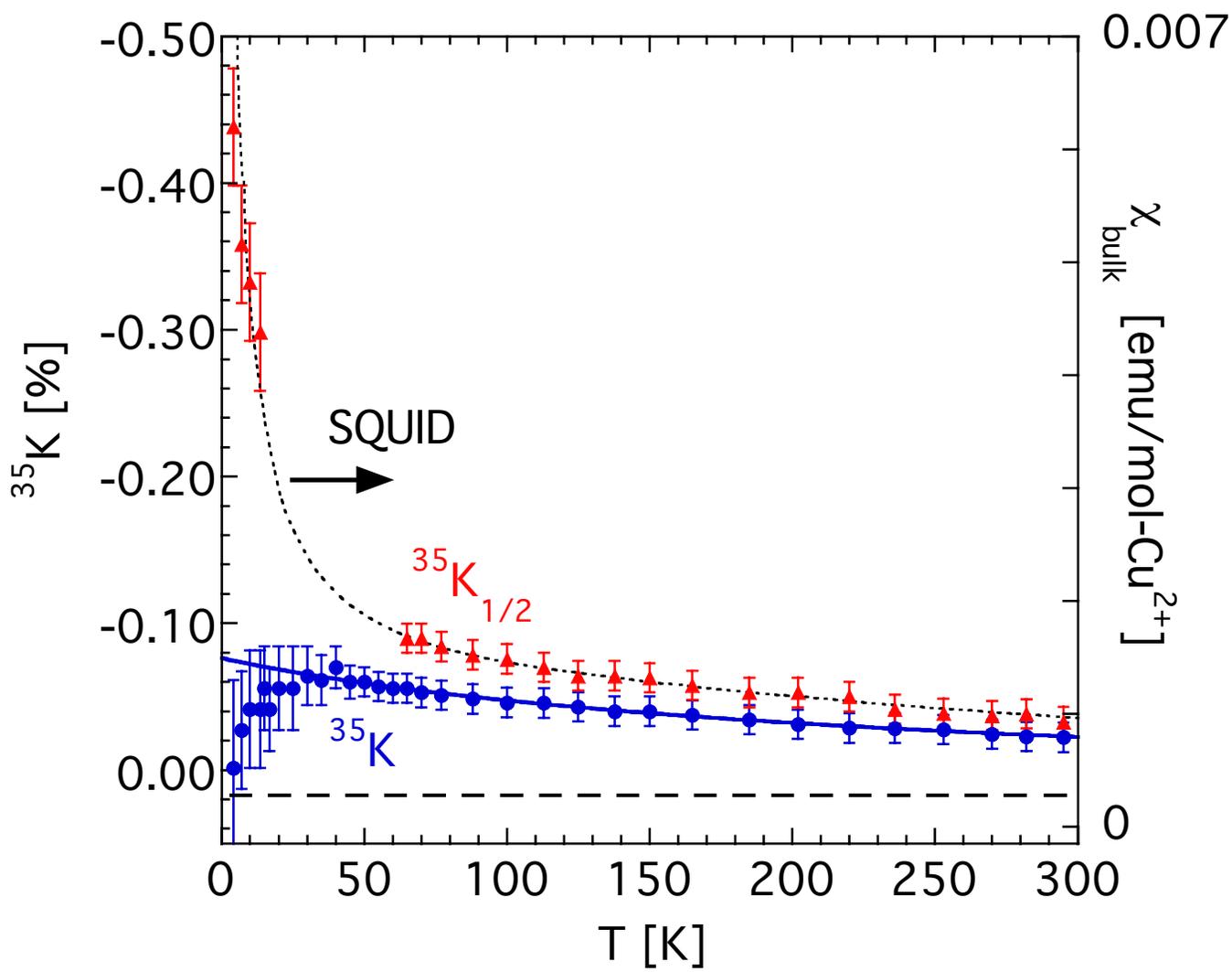
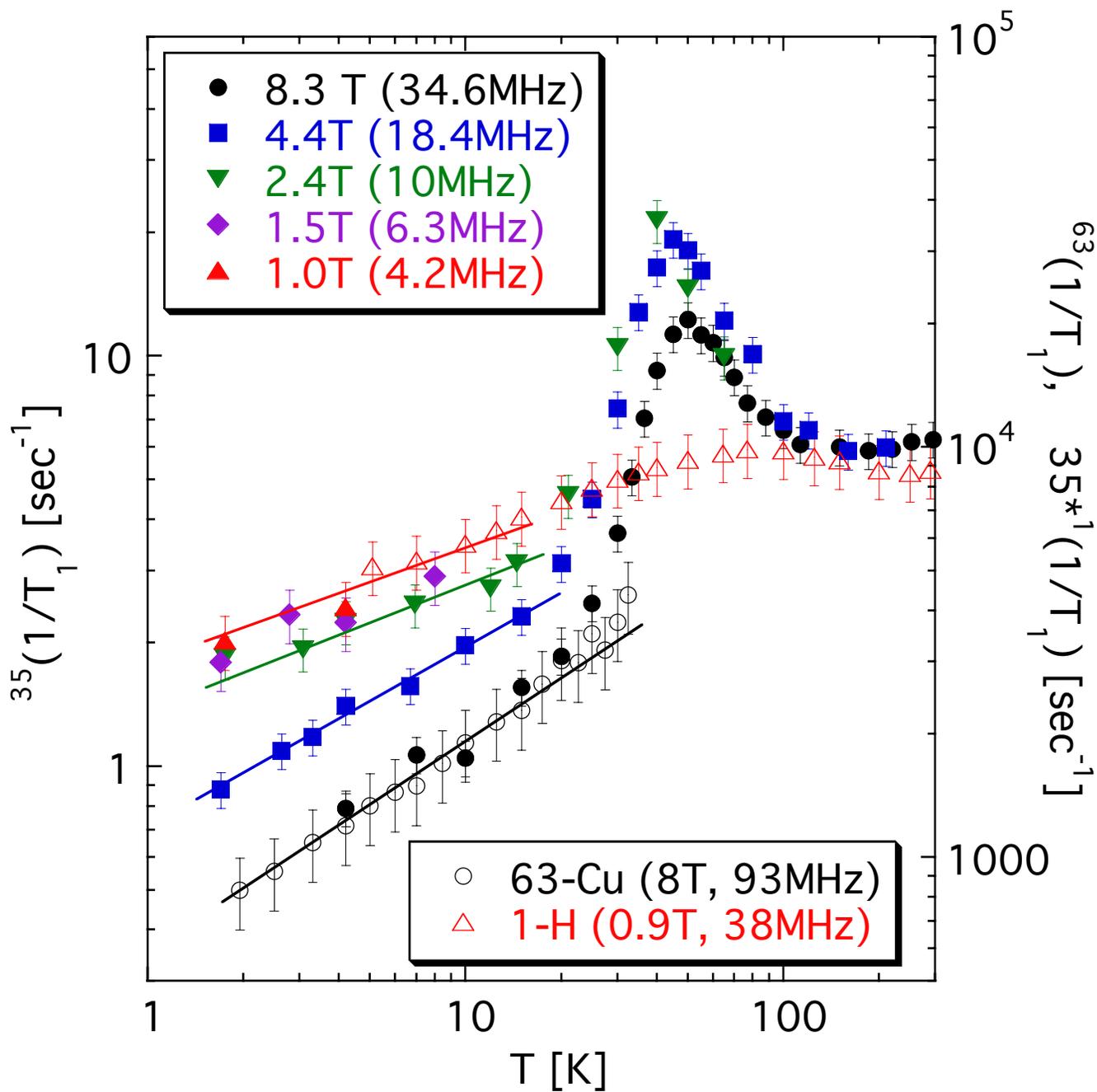
\end{document}